\documentstyle[12pt,epsf]{article}
\newcommand{\abstracts}[1]{{
\centering{\begin{minipage}{12.2truecm}
\normalsize\baselineskip=15pt
\centerline{\footnotesize ABSTRACT}\vspace*{0.3cm}
\parindent=20pt #1
\end{minipage}}\par}}
\newcommand{\LL}{{I\!\!  L}}

\newcommand{\eq}[1]{(\ref{#1})}

\newcommand{\beq}{\begin{equation}}
\newcommand{\eeq}{\end{equation}}
\newcommand{\beqn}{\begin{eqnarray}}
\newcommand{\eeqn}{\end{eqnarray}}

\newcommand{\cD}{{\cal D}}

\def\cC{{\cal C}}

\newcommand{\cZ}{{\cal Z}}

\def\dd{{\rm d}}
\newcommand{\Z}{{Z \!\!\! Z}}
\newcommand{\CK}[1]{\mbox{\scriptsize c}_{\mbox{$\scriptstyle #1$}}}
\newcommand{\nsum}[2]{\sum_{ #1(\CK{#2}) \in \Z }}

\newcommand{\nddsum}[2]{\sum_{\stackrel{\scriptstyle \dual #1(\dual\CK{#2})
\in \Z} {\delta \dual #1=0}}}
\newcommand{\dual}{\mbox{}^{\ast}}
\newcommand{\bbbz}{{\mathchoice {\hbox{$\sf\textstyle Z\kern-0.4em Z$}}
{\hbox{$\sf\textstyle Z\kern-0.4em Z$}}
{\hbox{$\sf\scriptstyle Z\kern-0.3em Z$}}
{\hbox{$\sf\scriptscriptstyle Z\kern-0.2em Z$}}}}
\newcommand{\intpiD}{\int\limits_{-\pi}^{+\pi} {\cD}}
\newcommand{\intinfD}{\int\limits_{-\infty}^{+\infty} {\cD}}
\newcommand{\expb}[1]{\exp\left\{ #1 \right\} }

\def\NP{ Nucl.~Phys.}

\def\PL{ Phys.~Lett.}
\def\PRL{ Phys.~Rev.~Lett.}

\begin{document}
~\vspace{-1.5cm}
\begin{flushright}
{\large ITEP-TH-16/98}\\
{\large MPI/PhT/98-24}\\
{\large March 1998}\\
\end{flushright}
\vspace{1.5cm}

\begin{center}

{\baselineskip=16pt
{\Large \bf Strings and Aharonov--Bohm Effect
\vspace{2mm}

in Abelian Higgs Model}\\

\vspace{1cm}

{\large M.N.~Chernodub\footnote{E-mail address:
chernodub@vxitep.itep.ru}, M.I.~Polikarpov\footnote{E-mail address:
polykarp@vxitep.itep.ru},
A.I.~Veselov and M.A.~Zubkov}\\

\vspace{.5cm}
{ \it

ITEP, B.Cheremushkinskaya 25, Moscow, 117259, Russia

}
}
\end{center}

\vspace{1cm}

\abstracts{We investigate numerically the properties of the Abrikosov
  $\!\!$--Nielsen--Olesen strings in $4D$ abelian Higgs model. The
  fractal dimension $D_f$ of the vortex strings was found to be large
  in the Coulomb phase and it is close to 2 in the Higgs phase.  We
  also show that the Wilson loop for non-integer charges is correlated
  with the linking number of the vortex string world sheets and the
  test particle world trajectory. We find that this topological
  (Aharonov--Bohm) interaction gives the main contribution to the
  Wilson loop quantum average for non--integer test charges in the
  vicinity of the Coulomb--Higgs phase transition.}

\newpage

\setcounter{footnote}{0}
\renewcommand{\thefootnote}{\alph{footnote}}

\section{Introduction}

The abelian Higgs model in $3+1$ dimensions possesses the vortex
topological defects called Abrikosov--Nielsen--Olesen (ANO) strings
\cite{AbNiOl}. These strings carry quantized magnetic flux and
therefore they interact with the charged particles through the analog
\cite{ABE} of the Aharonov--Bohm (AB) effect \cite{AhBo59}. The AB
effect has the topological origin: the corresponding interaction is
proportional to the linking number of the string world sheet and the
particle world trajectories. Due to AB interaction of the ANO strings
and charged particles the ANO strings may induce a long--range
interaction between the particles~\cite{GuPoCh96}. To study the role
of the AB effect in the particle--anti-particle potential we calculate
numerically in lattice Abelian Higgs model the correlations of the
Wilson loop on the contour $\cC$ and the linking number of the ANO
strings with this contour. We find that in the vicinity of the
Coulomb--Higgs phase transition the AB interaction provides the
dominant contribution to the Wilson loop.

Another aim of the paper is to study the fractal properties of the
ANO strings. We measure the fractal dimensionality $D_f$ of the ANO
strings and we find that $D_f$ is close to 2 in the Higgs phase and
it is much larger in the Coulomb phase. This fact is in agreement
with the observation \cite{Ch96} that the vortex strings are condensed
in the Coulomb phase and they are dilute in the Higgs phase.

In Section~2 we show the ANO strings representation for the partition
function of the $4D$ abelian Higgs model with the Villain action and
we show how the AB effect emerges in the path--integral formalism.
Section~3 is devoted to the discussion on the fractal properties of
the ANO strings and the role of the AB effect in the inter--particle
potential. Two quantities $R$ and $\cC$ which characterize the
contribution of the AB interaction to the inter--particle potentials
are suggested. These quantities and the fractal dimension of the ANO
strings are calculated numerically and the results are presented in
Section~4.

\section{Abrikosov--Nielsen--Olesen Strings and Aha\-ro\-nov--Bohm
Effect on the Lattice}

The partition function of $4D$ non--compact abelian Higgs model on
the lattice is\footnote{We are using the formalism of the differential
  forms on the lattice, see \cite{PoWiZu93,BeJo82} for the brief
  introduction.}:
\beq
  \cZ = \intinfD A \intpiD \varphi \nsum{l}{1} \expb{
  - \beta \|\dd A\|^2
  - \gamma \| \dd\varphi + 2\pi l + A \|^2}\,, \label{NC}\\
\eeq
where $A$ is the non--compact gauge field, $\varphi$ is the phase of
the Higgs field and $l$ is the integer--valued one--form. For
simplicity we consider the limit of the infinite Higgs boson mass
(the radial part of the Higgs field is frozen) and we use the
Villain form of the action.

One can rewrite \cite{PoWiZu93} integral \eq{NC} as the sum over
the closed string trajectories using the analogue of
Berezinski--Kosterlitz--Thauless (BKT) transformation \cite{BKT}:
\beq
  \cZ \propto \cZ^{BKT} = const. \, \nddsum{\sigma}{2}
  \expb{ - 4 \pi^2 \gamma
  \left(\dual \sigma, {(\Delta + m^2)}^{-1} \dual \sigma \right)}
  \,, \label{TD}
\eeq
where $m^2 = \gamma \slash \beta$ is the classical mass of the
vector boson $A$. Closed surfaces $\dual \sigma$ on the dual lattice
represent the string world sheets. It can be easily seen from
eq.\eq{TD} that the surfaces $\dual \sigma$ interact with each other
through the Yukawa forces.

Consider the quantum average of the Wilson loop
$W_{1 \over N}(\cC) = \exp\Bigl\{ i \frac{1}{N} (A,j_\cC)\Bigr\}$
for the particle with the charge $1 \slash N$. Using the BKT
transformation we get the following formula~\cite{PoWiZu93,Trento}:
\beqn
 {<W_{1 \over N}(\cC)>} = \frac{1}{\cZ^{BKT}} \nddsum{\sigma}{2}
 \exp\biggl\{- 4\pi^2\gamma \left(\dual \sigma,{\left(\Delta +
 m^2\right)}^{-1}\dual \sigma \right) \nonumber\\
 - \frac{1}{4 \beta N^2}( j_\cC,(\Delta + m^2)^{-1} j_\cC) + 2 \pi
 i \frac{1}{N} \left(\dual j_\cC,{\left(\Delta + m^2\right)}^{-1} \dd
 \dual \sigma \right) - 2 \pi i \frac{1}{N} \LL\left(\dual \sigma,
 j_\cC\right)\biggr\}\,.
 \label{wl}
\eeqn
The first three terms in this expression are the short--range Yukawa
forces between defects and particles. The last long--range term has
the topological origin: $\LL (\dual \sigma,j_\cC)$ is the linking
number between the world trajectories of the string world sheets $\dual
\sigma$ and the contour which define the Wilson loop,~$j_\cC$:
\beq
\LL(\dual \sigma,j_\cC) = (\dual j_\cC, {\Delta}^{-1} \dd \dual
\sigma) = (\dual j_\cC, \dual v)\,;
\,\, \dual \sigma = \delta \dual v\,.
\label{Ll}
\eeq
In four dimensions the trajectory of the string
$\dual \sigma$ is a closed surface and the linking number $\LL$ is
equal to the number of points at which the loop $j_\cC$ intersects the
three dimensional volume $\dual v$ bounded by the surface $\dual
\sigma$. The equation \eq{Ll} is the lattice analogue of the 4D Gauss
formula for the linking number. The strings can be considered as the
solenoids with magnetic flux which scatter the electrically charged
particles. After the scattering the wave--function of the charged
particle acquire an additional phase which is observable. This shift
in phase is the AB effect in the field
theory~\cite{ABE,PoWiZu93,Trento}.

\section{Fractal Dimension of Strings and Parameters of Topological
Interaction}

The abelian Higgs model with non-compact gauge field has Coulomb and
Higgs phases. The strings are dilute in the Higgs phase and they are
condensed in the Coulomb phase~\cite{Ch96}.
The condensed phase of the ANO strings is characterized by large
string clusters. One of the characteristics of the string
network is the fractal dimensionality $D_f$ which is defined as
follows. Let us denote the number of the plaquettes (links) of the
string world--sheet as $B$ (L). Each plaquette $P$ contributes the
value $|\dual \sigma_P|$ to the quantity $B$ and each link $l$
contributes to the quantity $L$ the value $\max_{P \ni l}
|\dual \sigma_P|$ if the link $l$ belongs to the string world sheets
or zero otherwise.  We define the fractal dimension as
\beq
D_f = 1+2 B/L\, . \label{Df}
\eeq
One can check, that if the
closed string has no selfintersections, then $D_f =2$. If the string
form 3-- or 4-- dimensional clusters, then $D_f$ equals to $3$ or $4$,
respectively. We expect, that if the strings are not condensed, then
$D_f$ is close to $2$ since the number of the self--intersections is
small. If the strings are condensed, then $D_f > 2$.

In the previous Section the topological (AB) interaction
of the ANO strings with the fractionally charged particles was shown.
This interaction is proportional to the linking number of the
particle trajectory and the string world sheet. One of the parameters
which can show the role of the AB effect in the quantum averages of
the Wilson loops is the following. The contribution of the given
field configuration to the Wilson loop on the contour $\cC$
is\footnote{Here we consider the real part of $W(\cC)$  since the
imaginary part vanishes in the quantum average $<W(\cC)>$.}
$\cos\Bigl(\frac{1}{N} (A,j_\cC)\Bigr)$, while the contribution to
this Wilson loop from the AB effect extracted from the same
configuration is:  $\cos\Bigl(\frac{2 \pi}{N} \LL(\dual \sigma,j_\cC)
\Bigr)$. The quantitative parameter which measures the correlation
between these two contributions is:  \beqn C(\cC,N) & = &
<\cos\Bigl(\frac{1}{N} (A,j_\cC)\Bigr)
\cos\Bigl(\frac{2\pi}{N} \LL(\dual \sigma,j_\cC) \Bigr)> - \nonumber\\
& & <\cos\Bigl(\frac1N (A,j_\cC)\Bigr)> \, <
\cos\Bigl(\frac{2\pi}{N} \LL(\dual \sigma,j_\cC) \Bigr)>\,.
\label{Cor}
\eeqn

If correlator \eq{Cor} is not zero then the AB effect
contributes into the quantum average of the Wilson loop.
Then another question appears: can this topological
interaction provide a dominant contribution to the Wilson loop?
To check that we study in the next Section the difference
between the Wilson loop average and the average of the AB
contribution to the Wilson loop:
\beqn
R = <\cos\Bigl(\frac1N (A,j_\cC)\Bigr)>- <
\cos\Bigl(\frac{2 \pi}{N} \LL(\dual \sigma,j_\cC) \Bigr)>\,.
\label{R}
\eeqn

\section{Numerical Results}

In numerical calculations we study the abelian Higgs model with the
Wilson type of the action in the London limit:
\beqn S = \beta {|| \dd
  A||}^2 - \gamma \sum_{l \in links} \cos {(\dd \phi + A)}_l\,.
\nonumber
\eeqn
The ANO strings are extracted from the field configurations as
follows~\cite{KRK83,Ch96}:
\beqn
\dual \sigma = \dual \dd n\,, \quad
(\dd\varphi + A + 2\pi n) \in
(-\pi,\pi]\,,\quad n\in \Z\,.  \nonumber
\eeqn

The calculations has been performed on $16^4$ lattice in the gauge
$\dd \phi = 0$. For each value of the parameters $(\beta , \gamma)$
50-100 statistically independent configurations have been generated
by the usual Monte-Carlo method. The dependence of the fractal
dimension $D_f$ \eq{Df} on the parameters $\beta$ and $\gamma$ is
shown on Figure~1(a)\footnote{The analogous study of the fractal
dimension of the global strings in 4D $XY$ model was performed in
ref.\cite{Sglob}}. The fractal dimension of the ANO strings is
close to $2$ in the Higgs phase and it is larger in the Coulomb
phase. Thus the string network in the Coulomb phase consists of the
clusters with the large number of selfintersections. The transition
from the dense string phase to the phase of the dilute strings is
rather sharp, see Figure~1(b).

Figure~2 shows the correlator $C$ \eq{Cor} for $4 \times 4$ Wilson
loop and for the charge $Q=1 \slash 10$ ($W(\cC) = \exp\{ i Q
(A,j_\cC\}$).  Since $C > 0$ then the topological interaction
contributes to the test particle interaction.

In Figure~3 (a,b) the quantity $R$ defined by eq.\eq{R} is shown. We
calculated $R$ for various contours $\cC$ ($3\times 3$, $3\times 4$,
$4\times 4$, $4\times 5$, $5\times 5$) and for various charges of the
test particle ($Q = 1/2, 1/3, 1/4, 1/10 $). It is interesting that the
zeroes of $R$ lie in the vicinity of the line of the phase
transition\footnote{Note that $R \to 0$ for $\beta \to 0$ since the
gauge boson mass is infinite in this limit and the AB effect
disappears.}. The positions of the zeroes depends very slightly on
the charge $N$ and on the shape of the Wilson loop. The parameter $R$
is negative in the Higgs phase and is positive in the Coulomb phase.
Thus near the phase transition the expectation value of the Wilson
loop is due to the topological (AB) interaction.

We conclude that the Abrikosov--Nielsen--Olesen strings in $4D$
Abelian Higgs model form a fractal cluster in the Coulomb phase ($D_f
>2$); in the Higgs phase $D_f \sim 2$, thus in this phase in the
string model \eq{TD} the strings are not crumpled. The
Aharonov--Bohm interaction of the strings with charged particles
gives the dominant contribution to the Wilson loop quantum average
for non--integer test charges in the vicinity of the Coulomb--Higgs
phase transition. Thus in this region the Abelian Higgs model behaves
as a topological theory, we study the origin of this fact in the
subsequent publications.

\section*{Acknowledgments}

M.N.Ch. and M.I.P. acknowledge the kind hospitality of the
Theoretical Department of the Kanazawa University. M.I.P. is indebted
for the warm hospitality at the Max--Plank--Institut f\"ur Physik,
M\"unchen. The authors are grateful to F.V.~Gubarev, M.~Laine and
V.I.~Zakharov for useful discussions. This work was supported by the
grants INTAS-96-370, INTAS-RFBR-95-0681, RFBR-96-02-17230a and
RFBR-96-15-96740.

\newpage
\section*{Figures}

\begin{figure*}[tbh]
\begin{center}
\begin{tabular}{cc}
\hspace{-0.8cm}\epsfxsize=6.7cm\epsffile{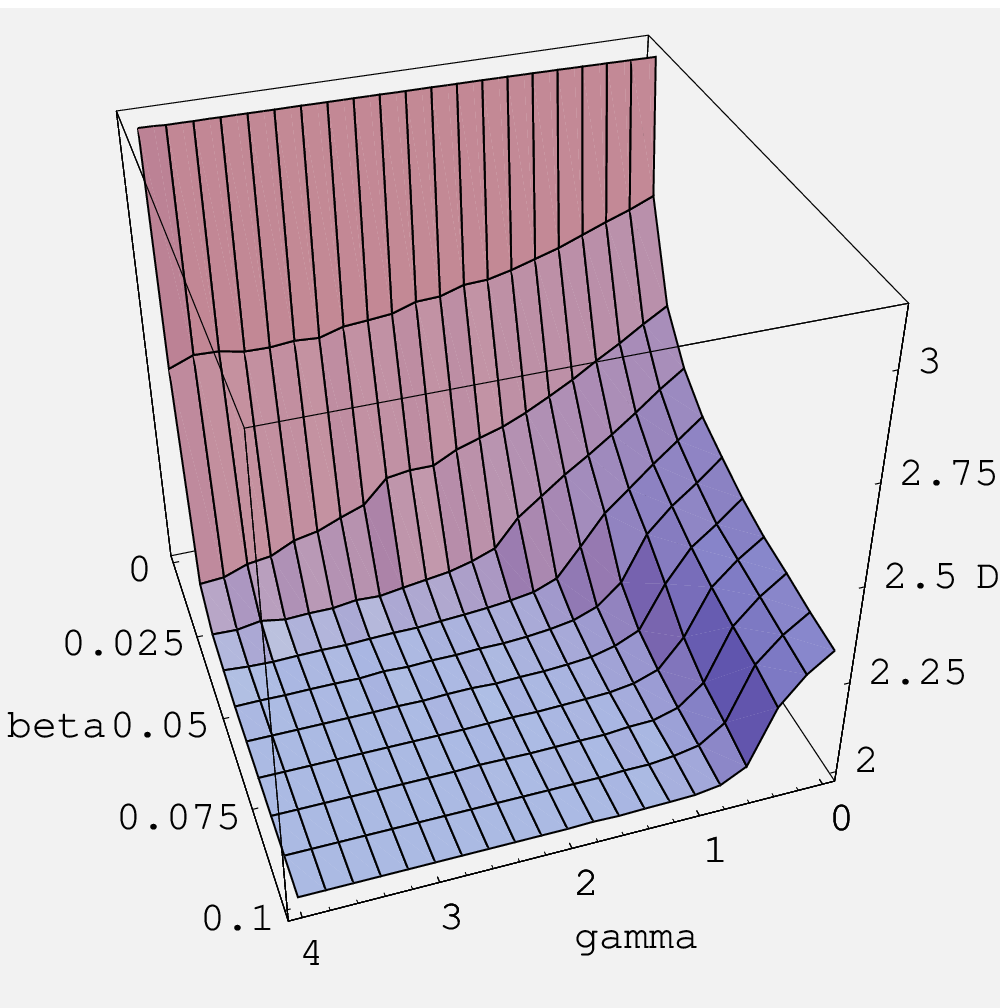} &
\hspace{0.8cm}\epsfxsize=6.7cm\epsffile{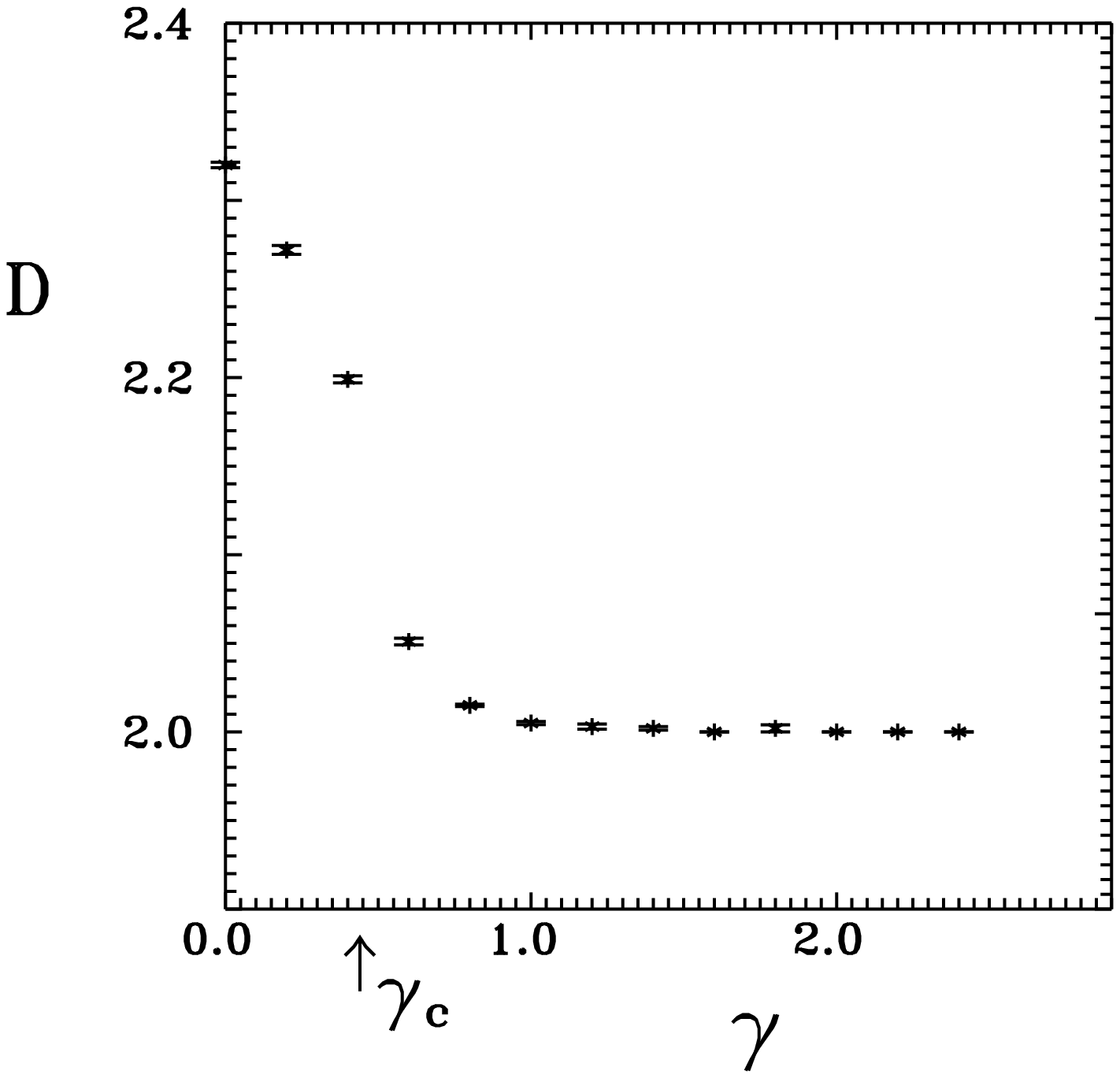} \\
(a) & (b)   \vspace{-0.3cm} \\ \\
\end{tabular}
\end{center}
\vspace{-0.5cm}
\caption{(a)The fractal dimension of the ANO strings, $D_f$, versus
$\beta$ and $\gamma$.  (b) The fractal dimension $D_f$, versus
$\gamma$ at $\beta = 0.1$; the phase transition from Coulomb phase
($\gamma < \gamma_c$) to Higgs phase ($\gamma > \gamma_c$) is at
$\gamma=\gamma_c\approx 0.48$.}
\end{figure*}

\begin{figure*}[t]
\begin{center}
\hspace{-0.8cm}\epsfxsize=6.7cm\epsffile{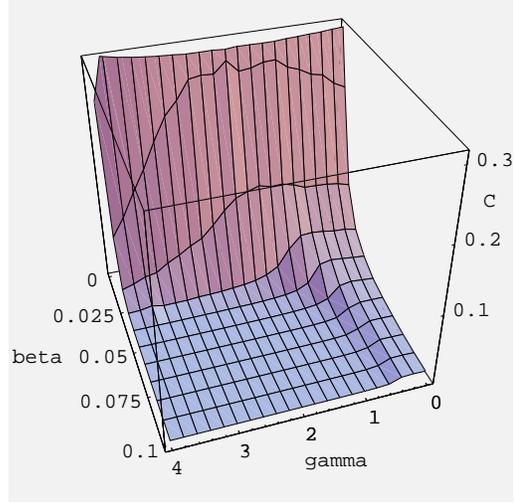}
\end{center}
\vspace{-0.5cm}
\caption{The correlator $C$ \eq{Cor}
for $4 \times 4$ Wilson loop, $Q = 1/10$.}
\end{figure*}

\begin{figure*}[t]
\begin{center}
\begin{tabular}{cc}
\hspace{-0.8cm}\epsfxsize=6.7cm\epsffile{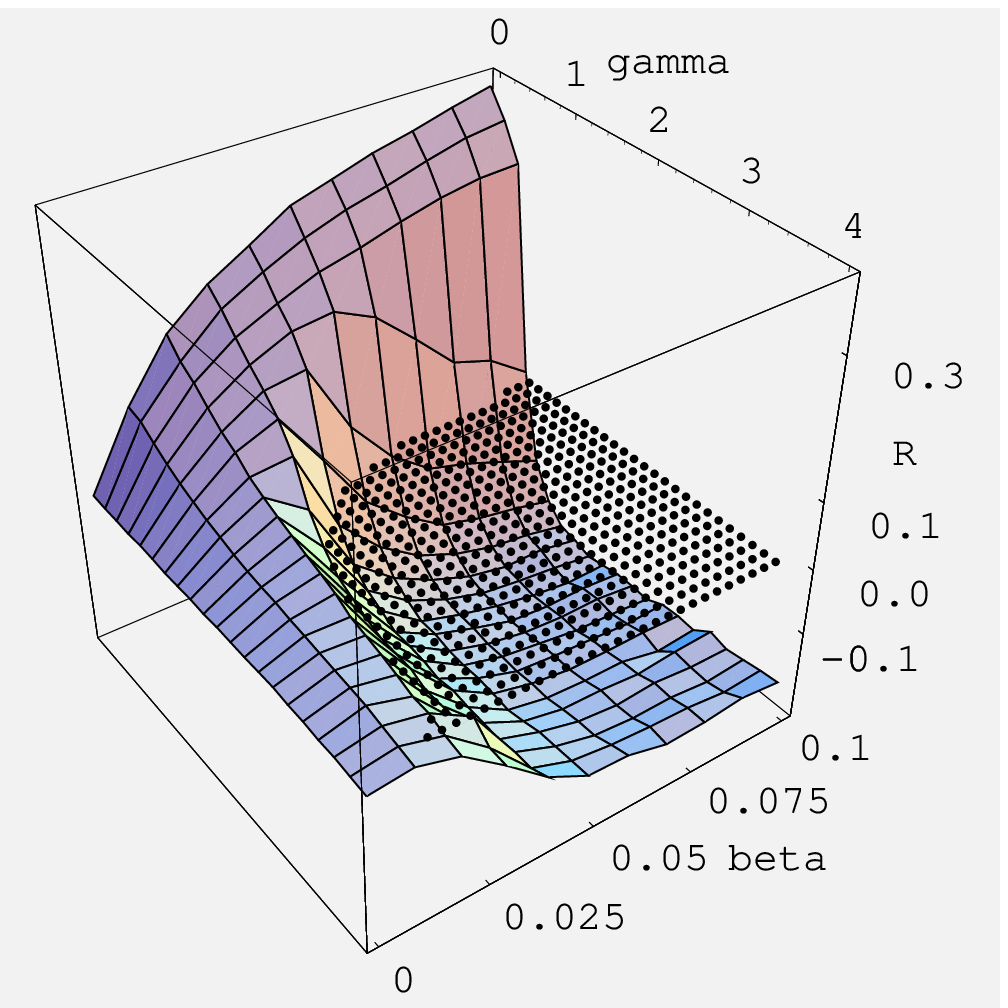} &
\hspace{0.8cm}\epsfxsize=6.7cm\epsffile{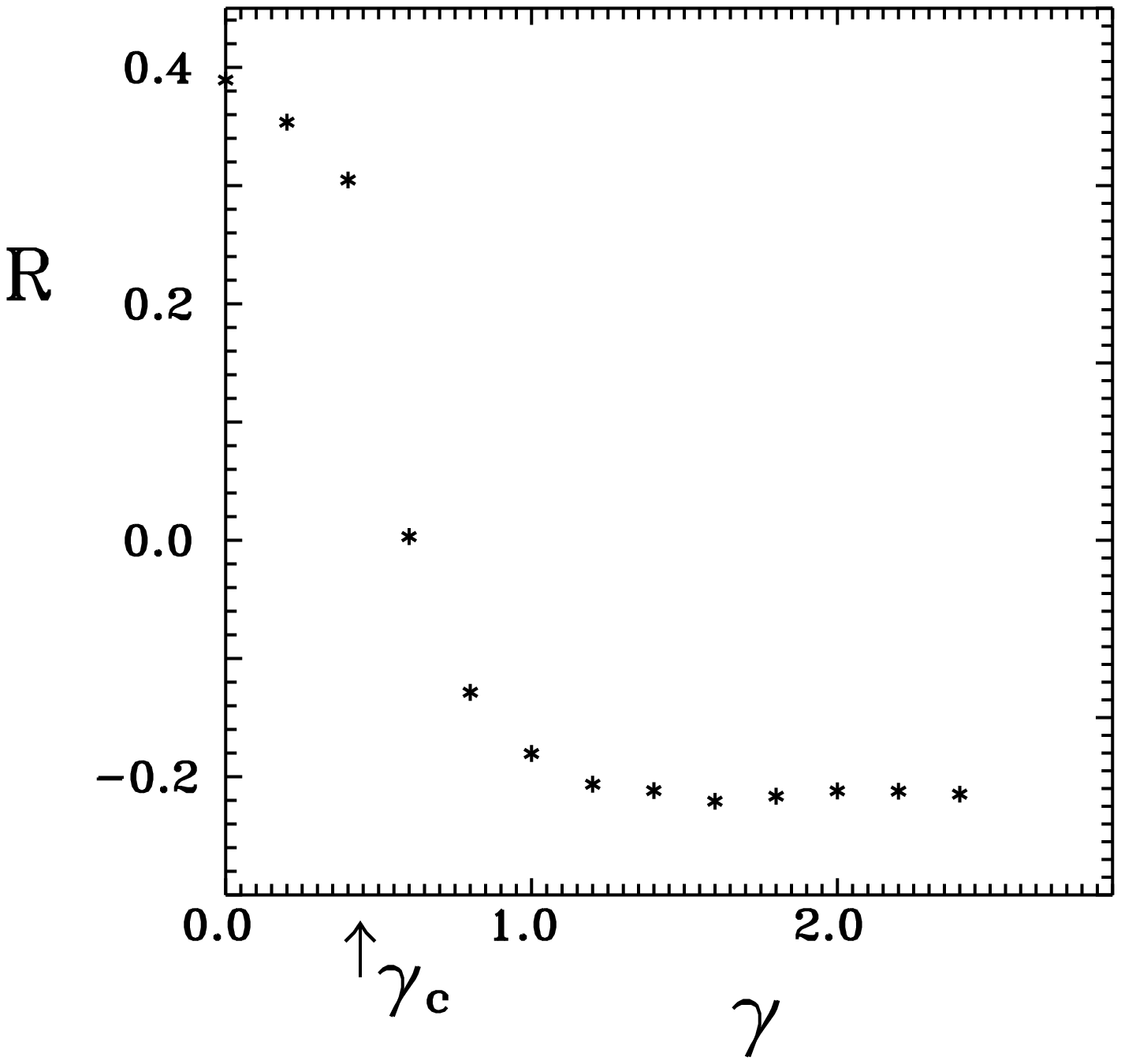} \\
(a) & (b)   \vspace{-0.3cm} \\ \\
\end{tabular}
\end{center}
\vspace{-0.5cm}
\caption{(a) The quantity $R$ \eq{R} for $4 \times 4$ Wilson loop, $Q
= 1/10$. The plane $R= 0$ is shown by dots. (b) The dependence of the
quantity $R$ on $\gamma$ at $\beta = 0.1$; the phase transition from
Coulomb phase ($\gamma < \gamma_c$) to Higgs phase ($\gamma >
\gamma_c$) is at $\gamma=\gamma_c\approx 0.48$.}
\end{figure*}


\newpage


\begin{thebibliography}{50}

\bibitem{AbNiOl}
A.A.~Abrikosov, {\it Sov. Phys. JETP}, {\bf 32} (1957) 1442;\\
H.B.~Nielsen and P.~Olesen, {\it \NP}, {\bf B61} (1973) 45.

\bibitem{ABE}
M.G.~Alford and F.~Wilczek, {\it \PRL}, {\bf 62} (1989) 1071;
L.M.~Kraus and F.~Wilczek, {\it \PRL}, {\bf 62} (1989) 1221;
M.G.~Alford, J.~March--Russel and F.~Wilczek, {\it \NP}, {\bf B337}
(1990) 695;
J.~Preskill and L.M.~Krauss, {\it \NP}, {\bf B341} (1990) 50;
E.T.~Akhmedov {et.al.}, {\it Phys.Rev.} {\bf D53} 2097 (1996).

\bibitem{AhBo59} Y. Aharonov and D. Bohm, {\it Phys.Rev.} {\bf 115}
(1959) 485.

\bibitem{GuPoCh96} F.A.~Bais, A.~Morozov, M.~de Wild~Propitius,
{\it \PRL} {\bf 71} (1993) 2383; M.N.~Chernodub, F.V.~Gubarev and
M.I.~Polikarpov, {\it JETP Lett}, {\bf 63} (1996) 516, {\tt
hep-lat/9607045}; \PL {\bf B416} (1998) 379; {\it
Nucl.Phys.Proc.Suppl.} {\bf 53} (1997) 581; {\tt hep-lat/9607045}.

\bibitem{Ch96} M.~Chavel, {\it Phys.~Lett.} {\bf B378} (1996) 227.

\bibitem{PoWiZu93}
M.I.~Polikarpov, U.-J.~Wiese and M.A.~Zubkov
{\it Phys. Lett.}, {\bf B 309} (1993) 133.

\bibitem{BeJo82}
P.~Becher and H.~Joos, {\it Z.~Phys.}, {\bf C15} (1982) 343.

\bibitem{BKT}
V.L.~Beresinskii, {\it {Sov. Phys. JETP}}, {\bf 32} (1970) 493;
J.M.~Kosterlitz and D.J.~Thouless, {\it J.~Phys.}, {\bf C6} (1973) 1181.


\bibitem{Trento}
M.N.~Chernodub and M.I.~Polikarpov,
{\it Proceedings of the International Workshop on Nonperturbative
Approaches to QCD, Trento, Italy, 10-29 Jul 1995},
{\tt hep-th/9510014}.

\bibitem{KRK83} J.~Ranft, J.~Kripfganz and G.~Ranft, {\it Phys.Rev.}
{\bf D28} (1983) 360.

\bibitem{Sglob} A.K.~Bukenov {\it et.al.}, {\it Phys. At. Nucl.}
  {\bf 56} (1993) 122 ({\it Yad.  Fiz.}  {\bf 56} (1993) 214).

\end{thebibliography}
\end{document}